# Investigation on the Spreading Behaviour of Sand Powder Used in Binder Jet 3D Printing

*Yulun Xu[1], Lanzhou Ge[1], Wenguang Nan[1\*, 2]*

1. School of Mechanical and Power Engineering, Nanjing Tech University, Nanjing 211816, China

2. Faculty of Engineering and Physical Sciences, University of Leeds, Leeds LS2 9JT, UK

*Contact Email: nanwg@njtech.edu.cn*

**Abstract:** The spreading behaviour of cohesive sand powder is modelled by Discrete Element Method, and the spreadability and the mechanical jamming are focused. The empty patches and total particle volume of the spread layer are examined, followed by the analysis of the geometry force and jamming structure. The results show that several empty patches with different size and shapes could be observed within the spread layer along the spreading direction even when the gap height increases to $3.0D_{90}$. Large particles are more difficult to be spread onto the base due to jamming, although their size is smaller than the gap height. Size segregation of particles occurs before particles entering the gap between the blade and base. There are almost no particles on the smooth base when the gap height is small, due to the full-slip flow of particles. The difference of the spread layer and spreadability between the cases with rough and smooth base is reduced by the increase of the gap height. An interesting correlation between jamming effect and local defects (empty spaces) in the powder layer is identified. The resistance to particle rolling is important for the mechanical jamming reported in this work. The jammed particles with a larger size ratio tend to be more stable.

# 1 Introduction

In powder-based Additive Manufacturing (AM), especially employing the binder jet 3D printing technology, the utilisation of cohesive powders is commonplace. However, these powders present challenges during the spreading process, and finally impact the quality of the final manufactured parts [1-3]. In the process of binder-jet sand printing, the sand particles are in angular shapes, and they are usually mixed with binder before spreading, making the powder very cohesive. The sand powder is spread by a blade spreader to form a thin and dense layer over a build work surface, and then another kind of binder is selectively spray onto the spread layer with locations corresponding to the 2D slice plane of the part. This process is repeated by lowering the build platform until the part is finished. However, the strong cohesion of powder and the angular shape of particles make the spreadability and the quality of sand layer not easily to be well controlled and thus induce quality problems for the final product.

The mechanism of powder spreading for thin layers in additive manufacturing has been explored by several researchers [4-13], including the effects of the spreading conditions and material properties on the quality of the spread layer. For example, Phua et al. [7] explored the effects of recoater geometry and speed on the granular convection and size segregation in powder spreading process. Nan et al. [14] and Ghadiri et al. [15] defined the spreadability of AM powder as ability of powder to be uniformly spread through a constriction to form a thin and dense layer without any defects such as empty patches and particle agglomerations. However, diversity of powder properties and sensitivity to spreading conditions make the prediction of spreadability challenging, but highly desirable, especially in binder jet 3D sand printing. Snow et al. [16] attempted to use four metrics to evaluate spreadability of metal powder through experiments, including the powder coverage percentage of substrate, the rate of powder deposition, the average avalanching angle of the powder heap, and the rate of change of the avalanching angle. Nan et al. [17] experimentally investigated the effect of particle properties on the spreadability of metal powder, and the results showed that excellent spreadability could only be obtained in a specified range of flowability, where the powder should be neither extremely cohesive nor excessively free-flowing with little frictional resistance. Lupo et al. [18] experimentally analysed the surface roughness of the spread layer of four kinds of polymer powders and found that the flowability and spreadability of these powders were not consistent. Shaheen et al. [19] explored the influence of particle material and process parameters on the spreadability, and found that irregular particles, rough particle surfaces and high interfacial cohesion mostly resulted in poor spreadability. Xu et al. [20] simulated the effects of the interfacial surface energy and rolling friction coefficient of single particle as well as the roughness of work surface on the spreadability. They concluded that powder spreadability was the combined effects of the shear action by the blade, the stagnation effect due to the rough base, particle jamming around the gap region, and the effect of powder flowability within the heap. They also clarified that jamming made powder spreadability different from powder flowability. Recently, Nan et al. [21] experimentally explored the characteristics of jamming for the blade spreading system, where metal powder and sand powder were used. Depending on the gap size normalised by particle $D_{90}$ diameter, particle interlocking caused by irregular particle shape, and particle cohesion, a regime map of jamming of granular flow through constriction was deduced in their work, including mechanical jamming, transition zone, cohesion-induced jamming, and slug or full slip.

Following our previous research [14, 20-22], the spreadability and jamming of powder used in binder-jet sand printing is detailed investigated in this work. The spreading process with a blade spreader is simulated by using Discrete Element Method (DEM). The spreadability is quantified by the empty patches and total particle volume of the spread layer, and the characteristics of mechanical jamming are analysed. It is followed by the discussions on the relationship between the spreadability and mechanical jamming. This provides a further step towards better understanding of powder spreadability and jamming of granular flow.

## 2 Methods

The particle flow in the spreading process is modelled by Discrete Element Method (DEM) using Altair EDEM software package, in which particle motion is tracked individually by solving Newton's laws of motion [23, 24]. The contact interaction force of particle against particle/wall is described by Hertz-Mindlin model with JKR theory [25], in which the normal contact force is given as:

$$F_n = \frac{4E^* a^3}{3R^*} - \sqrt{8\pi \Gamma E^* a^3} \qquad (1)$$

where $\Gamma$ is the interfacial surface energy; $E^*$ is the equivalent Young's modulus; $R^*$ is the equivalent radius; $a$ is the contact radius. An example of the variation of the normal contact force with the normal overlap is shown in Fig. 1. More details and the information of the damping force and tangential contact force could be referred to Thorton [24]. To describe the resistance of non-spherical particles to rolling, the rolling friction model developed by Ai et al. [26] is used, given as:

$$\boldsymbol{M}_r = \boldsymbol{M}_r^k + \boldsymbol{M}_r^d \qquad (2)$$

where $\boldsymbol{M}_r$ is the total rolling resistance, including a non-viscous term $\boldsymbol{M}_r^k$ and a viscous term $\boldsymbol{M}_r^d$, which are given as:

$$\boldsymbol{M}_r^k = -k_r \boldsymbol{\theta}_r \qquad (3)$$

$$|\boldsymbol{M}_r^k| \leq \mu_r R_r F_n \qquad (4)$$

$$\boldsymbol{M}_r^d = -2\eta \sqrt{I_r k_r} \boldsymbol{\omega}_r \qquad (5)$$

where $\boldsymbol{\theta}_r$ is the relative rotation angle; $\boldsymbol{\omega}_r$ is the relative rotational velocity; $R_r = R^*$ is the equivalent rolling radius; $I_r$ is the equivalent moment of inertia for the rotational vibration mode about the contact point; $\mu_r$ is the coefficient of rolling friction; $\eta$ is the rolling viscous damping ratio; $k_r$ is the rolling stiffness, given as:

$$k_r = 3k_n \mu_r^2 R_r^2 \qquad (6)$$

where $k_n$ is the normal contact stiffness. It should be noted that the viscous term $\boldsymbol{M}_r^d$ is only applied when the magnitude of the non-viscous term $\boldsymbol{M}_r^k$ is below the limit as defined in Eq. (4). More information could be referred to Ai et al. [26].

The powder-based spreading system is consisted of a rectangular blade, a baseplate and a powder heap, as shown in the Fig. 1. The base is in the same material as the powder, while the blade is 316L stainless steel. Sand powder in binder-jet printing of sand moulds is used in this work. The particle size distribution (PSD) based on particle number is shown in Fig. 1, as characterised by Mastersizer 2000 (Malvern Panalytical Ltd., Malvern, UK). The number-based $D_{10}$, $D_{50}$, $D_{90}$ are 100 μm, 140 μm, 207 μm [21], respectively. Although the particle shape is angular as shown in the SEM image in Fig. 1, spherical particles are used here for simplification, and the effects of shape on particle dynamics are described by using the rolling friction model described above and a large value of rolling friction coefficient. Meanwhile, totally 13 different particle sizes are used in the simulation, with the size distribution same as the original ones shown in Fig. 1. The friction angle is measured by the sliding method, as detailed described in Nan et al. [14]. The averaged friction of particles against particle/base and blade is 0.65 and 0.44, respectively. The surface energy $\Gamma$ and rolling friction coefficient $\mu_r$ are calibrated through experiments by reproducing the same static response angle (42 deg.) using the funnel method, and similar behaviour of the heap during the spreading process. The physical and mechanical properties of particles used in the simulation are summarised in Table 1. If not specified, the interaction parameters of particle-wall are assumed to be the same as that of particle-particle. A powder heap is firstly formed in front of the blade, and then the blade is lifted vertically in 0.005 s to the specified height $\delta$ above the top surface of the base. Afterwards, the blade accelerates quickly in $X$ direction in 0.02 s to reach the spreading speed of $U$=80 mm/s, which is in the range of the ones used in real equipment, and then the blade moves forward at this constant speed, followed by powder heap spreading onto the base. The computational domain is periodic in the $Y$ direction with a width of 3 mm. Four gap heights are used, i.e. 300 μm, 400 μm, 500 μm, 600 μm, which are about 1.5$D_{90}$, 2.0$D_{90}$, 2.5$D_{90}$, 3.0$D_{90}$, respectively. To investigate the formation mechanism of spread layer and spreading dynamics of sand powder, two kinds of rigid base are used, i.e. rough base and smooth base, as shown in Fig. 1. The former is represented by overlapping geometries, with a roughness of about 3.2 μm. The latter is represented by an ideal plate. It should be noted that the base used here is different from the real case of the powder spreading process in binder-jet sand printing, in which the base is comprised of particles bonded to each other through binder, allowing in a compressibility to some extent.

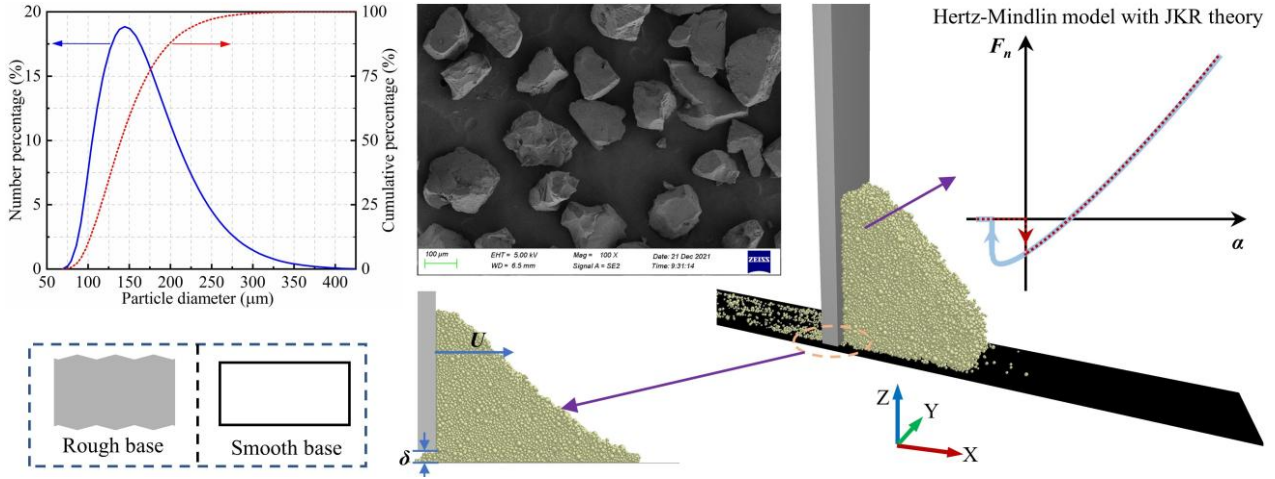

Fig. 1. Schematics of the spreading system in simulation and corresponding details.

Table 1. Physical and mechanical properties of particles used in simulation.

| Parameter | Value |
| --- | --- |
| Diameter, $D_{90}$ (μm) | 207 |
| Density, $\rho$ (kg/m$^3$) | 2650 |
| Young's modulus, $E$ (MPa) | 50 |
| Poisson's ratio, $v$ | 0.3 |
| Friction coefficient, $\mu$ | 0.65* |
| Restitution coefficient, $e$ | 0.7 |
| Rolling friction coefficient, $\mu_r$ | 0.8 |
| Interfacial surface energy, $\Gamma$ (mJ/m$^2$) | 8.0 |

* 0.44 is used for particle-blade interaction.

## 3 Results and discussions

### 3.1 Spread layer

Based on our previous work [14, 15, 20], spreadability is defined as the ability of powder to be uniformly spread through a constriction to form a thin and dense layer without any defects such as empty patches and particle agglomerations. Based on this concept, two kinds of metrics are used as the measurement of powder spreadability applied in thin layers for additive manufacturing, including empty patches and total particle volume of spread layer. Here, empty patches are referred to the area on the surface of the base which is not covered by any particles. Based on the snapshots of the spread layer with high resolution, total area of empty patches could be calculated through image analysis with the aid of ImageJ software package [27]. The total volume of particles within the spread layer $V_p$ is calculated and normalised by the width $W$ and length $L$ of the spread layer, given as [20]:

$$\phi = \frac{V_p}{LW} \quad (7)$$

Fig. 2 shows the snapshots of whole spread layer after the spreading process, and the corresponding area percentage of empty patches is shown in Fig. 3. In the case with rough base, a number of empty patches with different size and shapes are formed along the spreading direction. With the increase of gap height, fewer empty patches with smaller size could be observed within the spread layer. For example, the area percentage of empty patches decreases from 60% to 16% as the gap height increases from $1.5D_{90}$ to $3.0D_{90}$, as shown in Fig. 3. It should be noted that there are still empty patches even when the gap height is larger than $2.5D_{90}$. This is different to previous findings in Nan et al. [14], where the empty patches almost disappeared when the gap height increased to $2.5D_{90}$ and beyond. It may be due to the combined effects of strong cohesion and large rolling resistance of particles used in this work. Meanwhile, the data of the spreading experiment of sand powder carried out by Nan et al. [21] is also included in Fig. 2, which validates the simulation results to some extent.

Compared to the rough base, there are more empty patches at the same gap height for the cases with smooth base. There are almost no particles deposited on the base at gap height of $1.5D_{90}$ and $2.0D_{90}$, indicating the occurrence of a full-slip flow pattern of particles. Under this condition, the base has less ability to scrape particles out of the powder heap due to the lack of surface roughness, and thus powder heap tends to fully slip on the base. The scattered particles on the far left of the base are due to the destabilisation effect of the initially sudden moving of blade. It also suggests that the surface roughness has positive effects on the improvement of spreadability, due to the reduction of the probability of slug (full slip) flow pattern [21]. As the gap height increases to $2.5D_{90}$ and $3.0D_{90}$, the particles could be successfully spread onto base, although there are still some empty patches.

Correspondingly, the area percentage of empty patches decreases sharply from almost 100% to a value less than 50%.

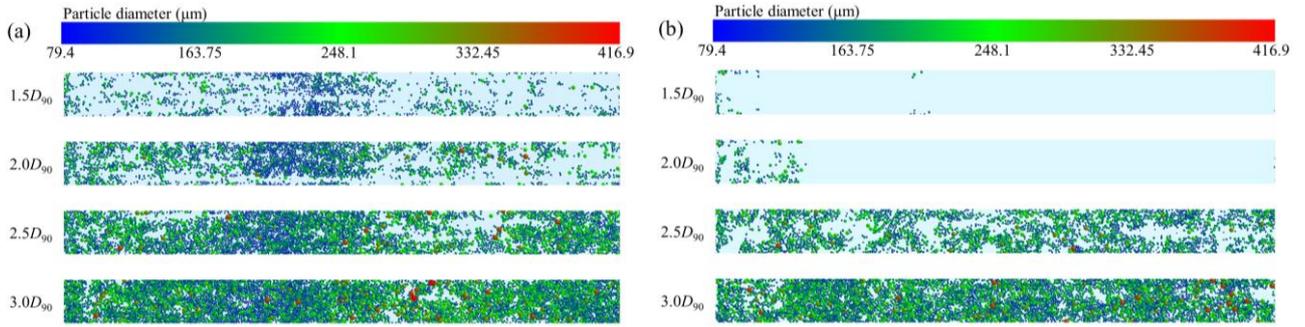

Fig. 2. Snapshots of deposited spread layer at different gap heights in the cases with (a) rough base and (b) smooth base, where particles are coloured based on particle diameter.

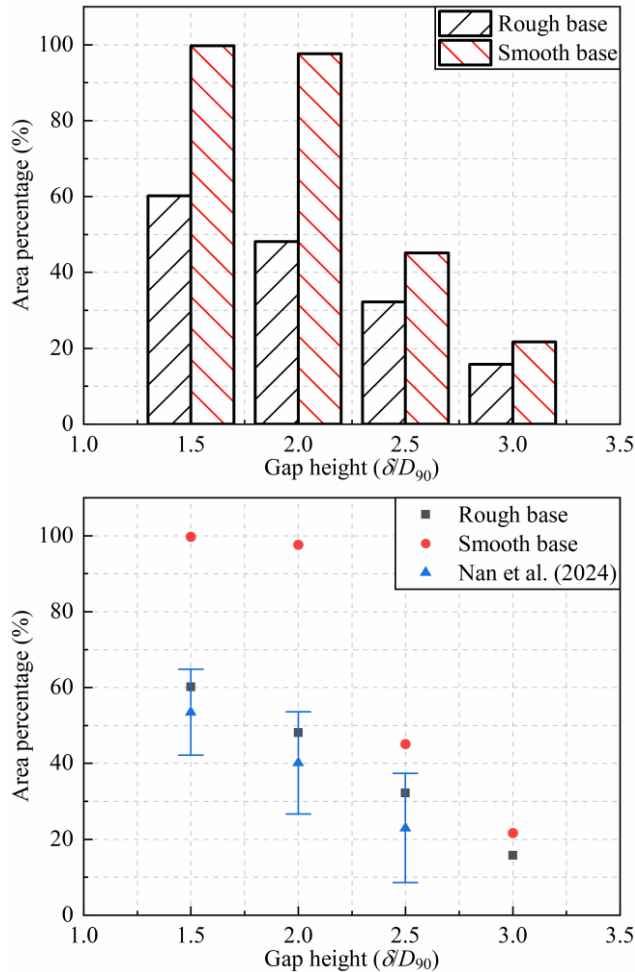

Fig. 3. Area percentage of empty patches at different gap heights in the cases with rough base and smooth base, where the experimental data from Ref. [21] is also included.

As shown in Fig. 2, the size distribution of deposited particles on the base varies with the gap height. In the case of rough base, with the increase of gap height, more and more particles with a large size could be observed within the spread layer. For example, almost all particles at the gap height of 1.5$D_{90}$ are in small size (blue to green colour), while largest particles (red colour) could be observed as the gap height increases to 3.0$D_{90}$. In the case with smooth base, largest particles are absent even when the gap height increases to 2.5$D_{90}$ and 3.0$D_{90}$. For a clear illustration, the size distribution of particles within the spread layer is analysed and shown in Fig. 4. As shown in Fig. 4(a), at gap height of 1.5$D_{90}$ (300 µm), the largest particle within the spread layer is 275.4 µm, which is expected as larger particles (i.e. 316.2, 363.1, 416.9 µm) could not pass through the gap due to size limit. With the increase of gap height, the size distribution curve moves right, indicating more and more particles with large size being spread onto the base. This is consistent with the observation in Fig. 2. The particle size distribution also becomes closer to the ones of the initial powder heap. It is interesting that at the gap height of 2.5$D_{90}$ (500 µm), the spread layer is in the absence of the largest particle (416.9 µm) although it is smaller than the gap height. It may be caused by jamming. As shown in Fig. 4(b), at both gap heights (2.5$D_{90}$ and 3.0$D_{90}$), there are almost no difference of the size distribution of particles within the spread layer between the cases with rough and smooth base. It indicates that the stagnation effects [20] due to base are minimised at large gap heights. Of course, all size distribution curves of particles within the spread layer are different to the ones of the initial powder heap, with more small particles and less large particles, indicating size segregation occurs before particles entering the gap between the blade and base. Several mechanisms may be responsible for this size segregation, such as percolation, shear-induced segregation, and also jamming reported in this work, which need to be further addressed in future.

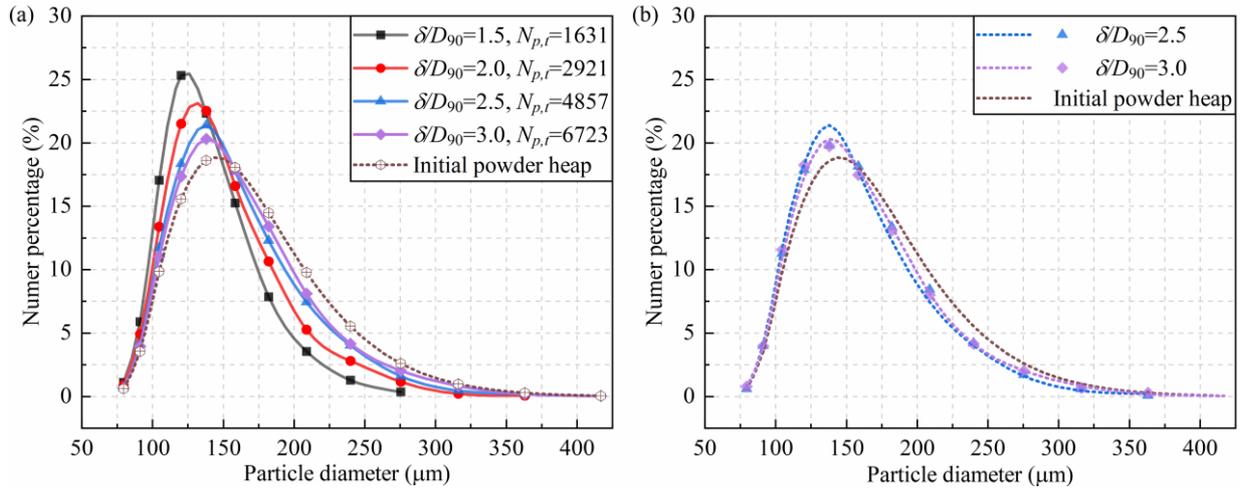

Fig. 4. Particle size distribution of spread layer: (a) the case with rough base, where the total number of particles is labelled; (b) comparison of the results between the cases with rough base (dotted line) and smooth base (solid symbol).

Fig. 5 shows the normalised total volume of particles within the spread layer. In the case with rough base, the total volume of particles for rough base increases almost linearly with the increase of gap height, as more particles could pass through the gap between the blade and base. Total particle volume of spread layer in the case of smooth base is much smaller than that for rough case at the same gap height, indicating again that the roughness of base is prone for better quality of spread layer. With the increase of gap height, the difference of total particle volume between the cases with rough base and smooth base is reduced. This is consistent with area percentage of empty patches shown in Fig. 3. It suggests that the gap height has more effects on the quality of spread layer than base roughness. In the case with smooth base, the curve of area percentage increases sharply when the gap height increases from $2.0D_{90}$ to $2.5D_{90}$. This is due to the sudden disappearance of slip flow pattern under the effect of large gap height.

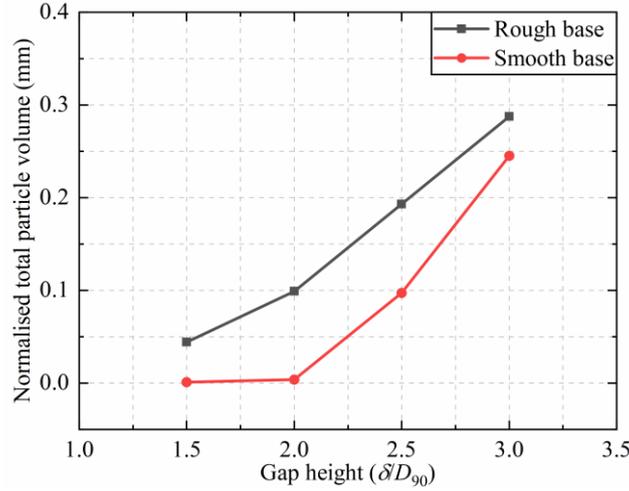

Fig. 5. Total particle volume of spread layer at different gap heights, which is normalised by the width and length of the spread layer shown in Eq. (7).

## 3.2 Mechanical jamming

Fig. 6 shows the variation of forces on the base and blade with time, where the force of base in vertical direction (Z direction) and the force of blade in spreading direction (X direction) are included. As shown in the Fig. 6(a), the trends of $F_{z,base}$ (solid line) and $F_{x,blade}$ (dotted line) are similar, although $F_{z,base}$ is a little larger than $F_{x,blade}$. As discussed in Nan et al. [21], the peak of forces could be used to represent the occurrence of mechanical jamming. It is interesting that with the increase of gap height, the number and value of the peaks first increase and then decrease quickly. It maybe due to that with the increase of the gap height, initially there is more space for particle entering into the gap region, and these abundant particles in contact with each other will enhance the particle jamming. However, with further increase of gap height, the wall effect would be much weakened, resulting in less occurrence of particle jamming. As shown in the Fig. 6(b), there are almost no fluctuations on the curves even when the gap heights are $1.5D_{90}$ and $2.0D_{90}$, indicating that the particles are mainly slipping on the base. As the gap height is increased to $2.5D_{90}$ and $3.0D_{90}$, several peaks could be observed, suggesting that the mechanical jamming could occur even when the base is smooth. The number and value of the peaks in the case with rough base are much larger than that of the case with smooth base at the same gap height. This suggests that rough base is more prone to generate mechanical jamming, and it could produce more violent jamming events than smooth base.

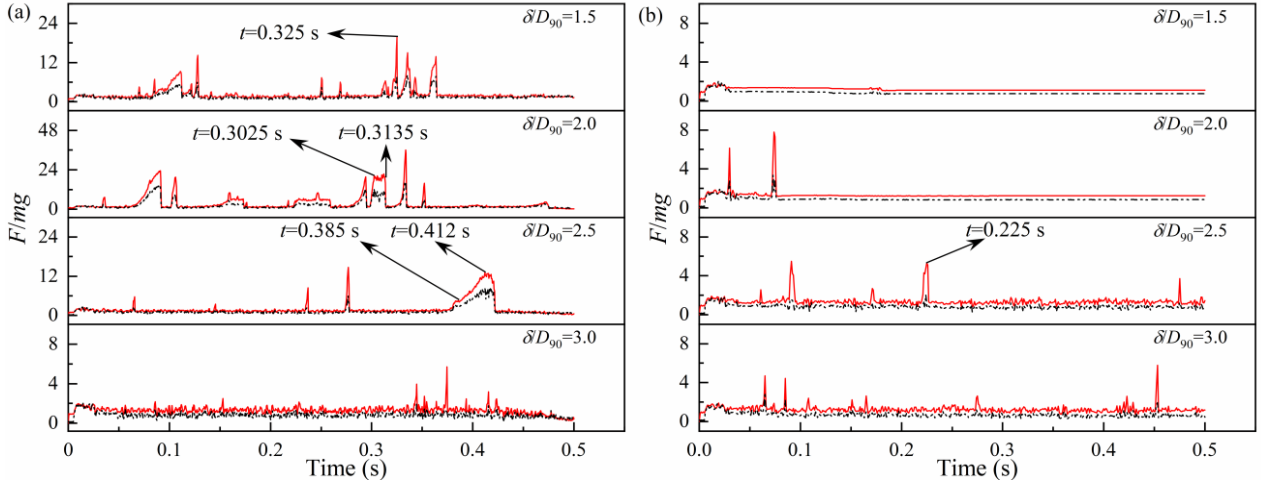

Fig. 6. Variation of $F_{z,base}$ (solid line) and $F_{x,blade}$ (dotted line) with time at different gap heights in the cases with (a) rough base and (b) smooth case.

Based on the characteristics of the peaks shown in Fig. 6, the jamming could be divided into two classes: short-jamming and long-jamming. The first refers to the transient force with large peaks, but this large force disappears quickly. Two examples are labelled in Fig. 6, i.e. $t=0.325$ s at the gap height of $1.5D_{90}$ with a rough base, and $t=0.225$ s at the gap height of $2.5D_{90}$ with a smooth base. The latter refers to the transient force with large peaks, and this large force could survive for a much longer period than that of the first. It is only observed at the gap height of $2.0D_{90}$ and $2.5D_{90}$ with a rough base. Two examples are labelled in Fig. 6, i.e. $t=0.3025$-$0.3135$ s at the gap height of $2.0D_{90}$, and $t=0.385$-$0.412$ s at the gap height of $2.5D_{90}$. The long-jamming has a lower frequency than that of short-jamming. For example, in the case of $\delta=2.5D_{90}$ with a rough base, there are only 1 long-jamming event while 4 short jamming events could be observed.

To illustrate the jamming structure, the force chain and jammed particles are shown in Fig. 7, where only the time steps labelled in Fig. 6 are included. For a clear illustration, only the jammed particles are shown while other particles are hidden, and the diameter of jammed particles is also measured. As shown in Fig. 7(a), a strong force chain is formed around the blade edge, and its direction is almost vertical. This is consistent with the spatial structure of jammed particles. Only 2 particles are jammed with a size ratio of 1.3, and the ratio of the size of largest particle to the gap height is 0.8. As shown in Fig. 7(b), it is interesting that in the case of smooth base, strong force chain could also be formed even at the gap height of 500 μm. Three particles are jammed in a chain structure with a vertical direction. It suggests that the occurrence of mechanical jamming is mainly controlled by the gap size instead of roughness of the base. The contact force of jammed particles in the case with rough base is about 5 times of that of smooth base, indicating that rough base could generate a stronger jamming.

For long-time jamming, to illustrate the evolution of jamming structure, the information at two time points is included. As shown in the Fig. 7(c)-(d), it is interesting that the jammed particles belong to the same particles during the whole jamming state. The fluctuation trend of force curve is consistent with the state of jammed particles. There are only 2 jammed particles, and the large particle is almost static while the small particle rolls around the large one. It indicates that the resistance to particle rolling is important for the mechanical jamming reported here. If the particle is free of rolling, the small particle would easily roll away from the bottom of large particle, thus, the jammed structure would be quickly damaged as the blade moves forward. Therefore, long-jamming is only observed in the case of rough base, as the roughness would promote enough resistance to the rolling of bottom particles. Meanwhile, the size ratio of the jammed particles is larger than that of short-jamming. For example, the size ratio is about 2 in Fig. 7(c) while it is 1.3 in Fig. 7(a). It indicates that the jammed particles with a large size ratio is more stable.

As summarised by Nan et al. [21], mechanical jamming has two adverse effects on the formation of spread layer: 1) the particle flow is transiently and locally halted, resulting in no particles on the base during the survival period of jamming; 2) large strain energy is stored during the survival period of jamming, and this energy would release suddenly and kick away the jammed particles once the jamming state is suddenly broken by the moving blade, resulting in particle colliding within the spread layer and significantly inducing non-uniform distribution of particles within the spread layer. Following this concept, the formation of empty patches maybe related to the occurrence of mechanical jamming. Here, the vertical force on the base is plotted against the snapshot of the spread layer, as shown in Fig. 8, where only the results at gap heights of $2.0D_{90}$ and $2.5D_{90}$ in the case of rough base are included for brevity. The location $x$ of the spread layer is linked to the time by $x=U\times t-th$, where *th* is the thickness of the blade. When the jamming events are severe (more and larger peaks of force), the corresponding spread layer has more empty patches. Especially, when the long-time jamming occurs, as boxed by a dashed box, a long strip could be observed at the corresponding locations within the spread layer. It indicates that empty patches could be induced by the mechanical jamming.

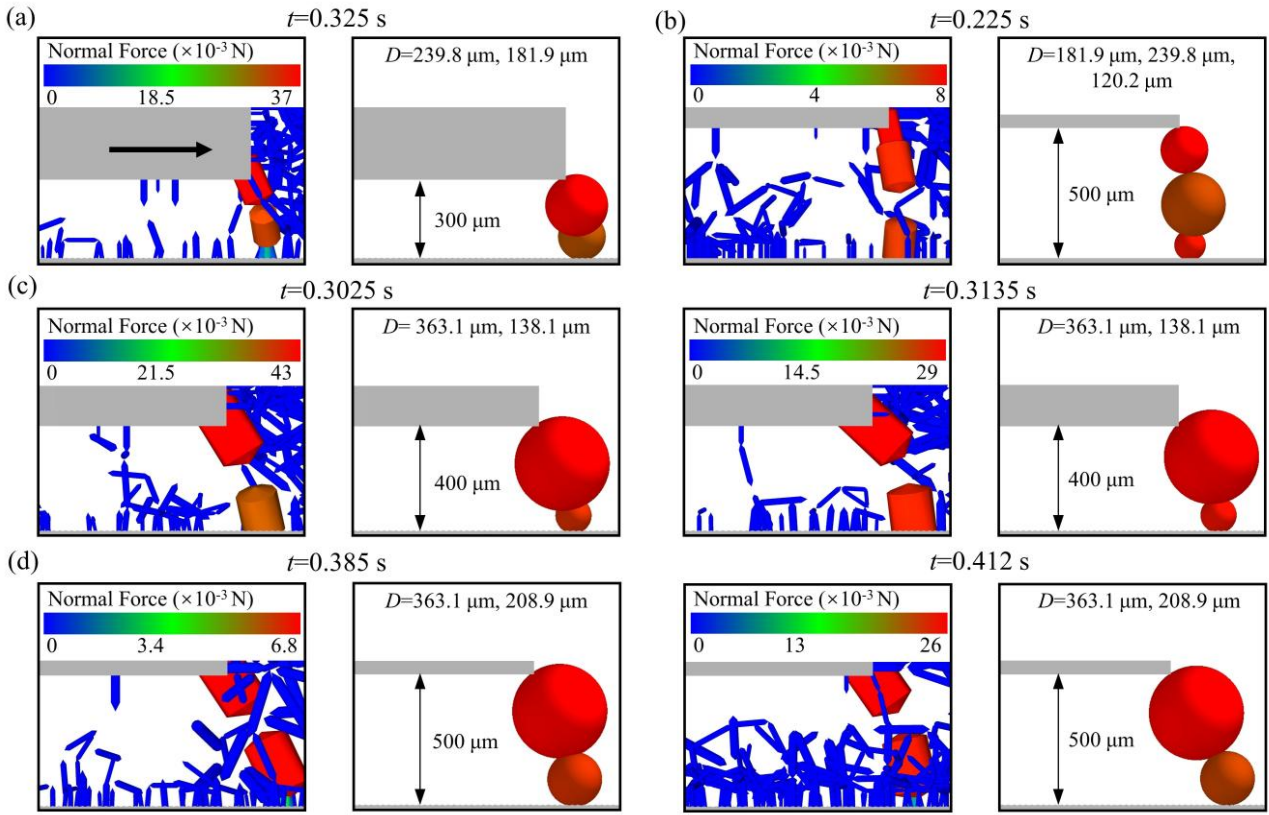

Fig. 7. Snapshots of the force chain and jammed particles: (a) gap height of $1.5D_{90}$ with rough base. (b) gap height of $2.5D_{90}$ with smooth base. (c) gap height of $2.0D_{90}$ with rough base. (d) gap height of $2.5D_{90}$ with rough base.

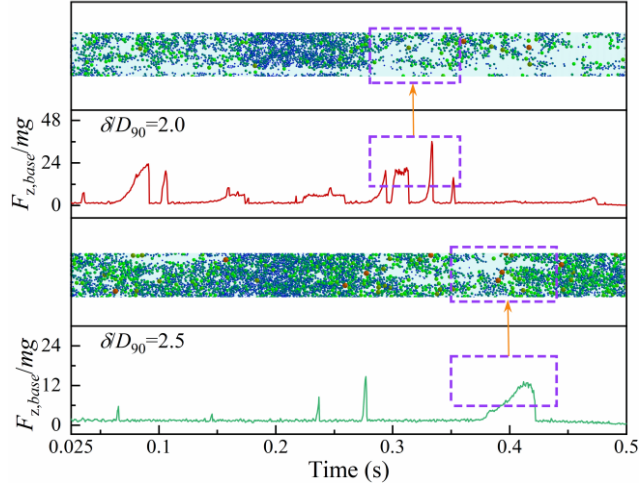

Fig. 8. Relationship between the vertical force on the base and empty patches within the spread layer at the gap height of 2.0$D_{90}$ and 2.5$D_{90}$ in the case with rough base.

**4 Conclusion**

The spreading behaviour of cohesive sand powder is simulated by using DEM, and the effects of gap heights and base roughness on the spreadability and mechanical jamming of particles are investigated. The empty patches and total particle volume are examined, followed by the analysis of the geometry force and jamming structure. Main results from the present study are summarised as follows:

1) Several empty patches with different size and shapes could be observed within the spread layer along the spreading direction. Empty patches still exist even when the gap height is 3.0$D_{90}$ due to strong cohesion and large rolling resistance of particles used in this work. Large particles are more difficult to be spread onto the base due to jamming, although their size is smaller than the gap height. Size segregation of particles occurs before particles entering the gap between the blade and base.

2) There are almost no particles on the smooth base when the gap height is small, due to the full-slip flow pattern of particles. Mechanical jamming and strong force chain could be formed even on the smooth base at large gap heights. The difference of the spread layer and spreadability between the cases with rough and smooth bases is reduced by the increase of the gap height.

3) Two kinds of mechanical jamming are observed depending on its survival period. The resistance to particle rolling is important for the mechanical jamming reported in this work. The jammed particles with a larger size ratio tend to be more stable. An interesting correlation between jamming effect and local defects (empty spaces) in the powder layer is identified.

**Acknowledgments**

The authors are grateful to the National Key Research and Development Program of China (Grant No. 2022YFB4602202) and National Natural Science Foundation of China (Grant No. 51806099). The second author is also thankful to Postgraduate Research & Practice Innovation Program of Jiangsu Province (Grant No. KYCX23_1440). The corresponding author is also thankful to Professor Mojtaba Ghadiri, University of Leeds, UK, for the inspiration on this work.

**Compliance with ethical standards**

Conflict of Interest: The authors declare that they have no conflict of interest that could have appeared to influence the work reported in this paper.

**References**

[1] B. Yang, Y. Wang, Y. Zhang, The 3-D spread of saltation sand over a flat bed surface in aeolian sand transport, Advanced Powder Technology, 20 (2009) 303-309.

[2] M. Upadhyay, T. Sivarupan, M. El Mansori, 3D printing for rapid sand casting—A review, Journal of Manufacturing Processes, 29 (2017) 211-220.

[3] T. Sivarupan, N. Balasubramani, P. Saxena, D. Nagarajan, M. El Mansori, K. Salonitis, M. Jolly, M.S. Dargusch, A review on the progress and challenges of binder jet 3D printing of sand moulds for advanced casting, Additive Manufacturing, 40 (2021).

[4] D. Nasato, T. Pöschel, Influence of particle shape in additive manufacturing: Discrete element simulations of polyamide 11 and polyamide 12, 36 (2020) 101421.

[5] D. Schiochet Nasato, H. Briesen, T. Pöschel, Influence of vibrating recoating mechanism for the deposition of powders in additive manufacturing: Discrete element simulations of polyamide 12, Additive Manufacturing, 48 (2021) 102248.

[6] D. Yao, X. An, H. Fu, H. Zhang, X. Yang, Q. Zou, K. Dong, Dynamic investigation on the powder spreading during selective laser melting additive manufacturing, Additive Manufacturing, 37 (2021) 101707.

[7] A. Phua, C. Doblin, P. Owen, C.H.J. Davies, G.W. Delaney, The effect of recoater geometry and speed on granular convection and size segregation in powder bed fusion, Powder Technology, 394 (2021) 632-644.

[8] D. Ruggi, M. Lupo, D. Sofia, C. Barrès, D. Barletta, M. Poletto, Flow properties of polymeric powders for selective laser sintering, Powder Technology, 370 (2020) 288-297.

[9] E.J.R. Parteli, T. Poschel, Particle-based simulation of powder application in additive manufacturing, Powder Technology, 288 (2016) 96-102.

[10] S. Haeri, Y. Wang, O. Ghita, J. Sun, Discrete element simulation and experimental study of powder spreading process

in additive manufacturing, Powder Technology, 306 (2016) 45-54.

[11] S. Haeri, Optimisation of blade type spreaders for powder bed preparation in Additive Manufacturing using DEM simulations, Powder Technology, 321 (2017) 94-104.

[12] P.S. Desai, A. Mehta, P.S.M. Dougherty, C.F. Higgs, A rheometry based calibration of a first-order DEM model to generate virtual avatars of metal Additive Manufacturing (AM) powders, Powder Technology, 342 (2019) 441-456.

[13] H. Chen, Q.S. Wei, S.F. Wen, Z.W. Li, Y.S. Shi, Flow behavior of powder particles in layering process of selective laser melting: Numerical modeling and experimental verification based on discrete element method, Int J Mach Tool Manu, 123 (2017) 146-159.

[14] W. Nan, M. Pasha, B. Tina, L. Alejandro, Z. Umair, N. Sadegh, M. Ghadiri, Jamming during particle spreading in additive manufacturing, Powder Technology, 338 (2018) 253–262.

[15] M. Ghadiri, M. Pasha, W. Nan, C. Hare, V. Vivacqua, U. Zafar, S. Nezamabadi, A. Lopez, M. Pasha, S. Nadimi, Cohesive Powder Flow: Trends and Challenges in Characterisation and Analysis, KONA Powder and Particle Journal, 37 (2020) 3-18.

[16] Z. Snow, R. Martukanitz, S. Joshi, On the development of powder spreadability metrics and feedstock requirements for powder bed fusion additive manufacturing, Additive Manufacturing, 28 (2019) 78-86.

[17] W. Nan, Y. Gu, Experimental investigation on the spreadability of cohesive and frictional powder, Advanced Powder Technology, 33 (2022) 103466.

[18] M. Lupo, S.Z. Ajabshir, D. Sofia, D. Barletta, M. Poletto, Experimental metrics of the powder layer quality in the selective laser sintering process, Powder Technology, 419 (2023) 118346.

[19] M.Y. Shaheen, A.R. Thornton, S. Luding, T. Weinhart, The influence of material and process parameters on powder spreading in additive manufacturing, Powder Technology, 383 (2021) 564-583.

[20] R. Xu, W. Nan, Analysis of the metrics and mechanism of powder spreadability in powder-based additive manufacturing, Additive Manufacturing, 71 (2023).

[21] W. Nan, L. Ge, W. Xuan, Y. Gu, Transient jamming of granular flow by blade spreading, Powder Technology, 431 (2024).

[22] W. Nan, M.A. Rahman, L. Ge, Z. Sun, Effect of plastic deformation on the spreadability of cohesive powder in the spreading process, Powder Technology, 425 (2023).

[23] P.A. Cundall, O.D.L. Strack, A discrete numerical model for granular assemblies, Geotechnique, 29 (1979) 47-65.

[24] C. Thornton, Granular Dynamics, Contact Mechanics and Particle System Simulations: A DEM Study, 2015.

[25] K.L. Johnson, K. Kendall, A.D. Roberts, D. Tabor, Surface energy and the contact of elastic solids, Proceedings of the Royal Society of London. A. Mathematical and Physical Sciences, 324 (1971) 301-313.

[26] J. Ai, J.-F. Chen, J.M. Rotter, J.Y. Ooi, Assessment of rolling resistance models in discrete element simulations, Powder Technology, 206 (2011) 269-282.

[27] J. Schindelin, I. Arganda-Carreras, E. Frise, V. Kaynig, M. Longair, T. Pietzsch, S. Preibisch, C. Rueden, S. Saalfeld, B. Schmid, J.Y. Tinevez, D.J. White, V. Hartenstein, K. Eliceiri, P. Tomancak, A. Cardona, Fiji: an open-source platform for biological-image analysis, Nat Methods, 9 (2012) 676-682.